\documentclass[conference]{IEEEtran}
\IEEEoverridecommandlockouts
\usepackage{comment}
\usepackage{array}
\usepackage{cite}
\usepackage{upgreek}
\usepackage{hyperref}
\usepackage{amsmath,amssymb,amsfonts}
\usepackage{algorithmic}
\usepackage{graphicx}
\usepackage{multirow}
\usepackage{textcomp}
\usepackage{xcolor}
\def\BibTeX{{\rm B\kern-.05em{\sc i\kern-.025em b}\kern-.08em
    T\kern-.1667em\lower.7ex\hbox{E}\kern-.125emX}}

\def\etc{\textit{etc}}

\newcommand{\RS}[1]{{\color{black}{#1}}}

\usepackage{fancyhdr}
\begin{document}
\title{A Weakly Supervised Approach to Emotion-change Prediction and Improved Mood Inference
\thanks{This research is partially funded by the Australian Government through the Australian Research Council’s Discovery Projects funding scheme (project DP190101294). \\
*Corresponding author: soujanya.narayana@canberra.edu.au}
}

    \author{\parbox{16cm}{\centering
    {\large Soujanya Narayana$^{1*}$, Ibrahim Radwan$^1$, Ravikiran Parameshwara$^1$, Iman Abbasnejad$^2$, Akshay Asthana$^2$, Ramanathan Subramanian$^1$ and Roland Goecke$^1$}\\
    {\normalsize
    $^1$Human-Centred Technology Research Centre, University of Canberra, Australia}\\
    $^2$Seeing Machines Ltd., Australia\\
    }}

\maketitle
\thispagestyle{fancy}

\begin{abstract}
Whilst a majority of affective computing research focuses on inferring emotions, examining mood or understanding the \textit{mood-emotion interplay} has received significantly less attention. Building on prior work, we (a) deduce and incorporate emotion-change ($\Delta$) information for inferring mood, without resorting to annotated labels, and (b) attempt mood prediction for long duration video clips, in alignment with the characterisation of mood. We generate the emotion-change ($\Delta$) labels via metric learning from a pre-trained Siamese Network, and use these in addition to mood labels for mood classification. Experiments evaluating \textit{unimodal} (training only using mood labels) vs \textit{multimodal} (training using mood plus $\Delta$ labels) models show that mood prediction benefits from the incorporation of emotion-change information, emphasising the importance of modelling the mood-emotion interplay for effective mood inference.
\end{abstract}


\section{Introduction}

\begin{figure*}[t]
\centering
\includegraphics[width=0.95\textwidth]{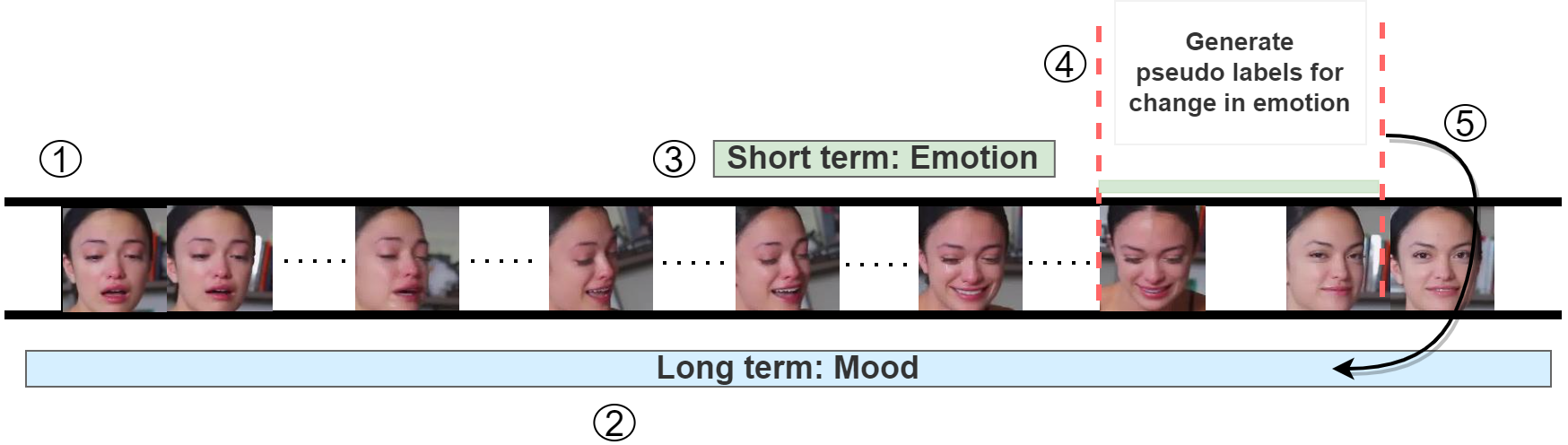}\vspace{-1mm}
\caption{\textbf{Study Overview:} (1) We consider the publicly available \textit{AffWild2}~\cite{kollias2019expression} video dataset for automated mood inference. (2) A mood label is derived for the video, aligning with the characterisation of mood as a long-term affective state. (3) Our study also seeks to automatically capture affective emotion-change ($\Delta$) information over shorter durations (time windows). (4) We generate pseudo-emotion-change ($\Delta$) labels via metric learning obviating the need for valence annotations. (5) We then incorporate the generated $\Delta$ labels to automatically infer the mood class as \textit{positive}, \textit{negative} or \textit{neutral}.}
\label{fig:overview}
\end{figure*}

Over the past two decades, there has been an enormous increase in the research on inferring affective states (characterised by emotions, moods, \etc.) from unimodal and multimodal data. Several studies emphasise on the importance of emotional regulation for the successful functioning of human mind~\cite{hudlicka1996review, desteno2013affective}, as they play an indispensable role in rational decision-making, perception, attention, and other diverse cognitive functions \cite{picard2000affective}. While the terms \emph{emotion} and \emph{mood} are often used synonymously, the two affective phenomena are distinct in terms of duration, intensity, attribute, and behavioural impact. \emph{Emotion} is a short-term affective state, lasting for at most a few minutes, and is typically elicited by a contextual event/stimulus. On the contrary, \emph{mood} is considered to be a long-term diffuse affective state lasting for hours, which may emerge without an apparent cause~\cite{scherer2005emotions}. 

Akin to emotions, mood has an impact on cognitive processes like human creativity, evaluative judgement, and memory retrieval, etc \cite{picard2000affective}. Mood is also known to generate cognitive bias and influence human emotion recognition~\cite{siemer2001mood}. Going further, mood disorders like depression and bipolar disorder result in the impairment of emotional processing abilities~\cite{panchal2019cognitive}, altered facial expression understanding~\cite{venn2006facial}, and olfactory perception~\cite{lombion2006odor}. Recently, the focus of numerous studies is on building emotionally-aware systems to better understand human behaviour, and facilitate enhanced human-computer interactions. Advancements in machine learning techniques have enabled automatic emotion recognition from unimodal and multimodal data, for instance, facial expressions~\cite{tarnowski2017emotion}, physiological signals \cite{Bilalpur17,Shukla17,Subramanian18}, videos~\cite{parameshwara2023examining}, \etc. Although substantial progress has been made in psychology to understand mood, negligible work has focused on computationally inferring mood. Furthermore, the psychology literature recognises an association between emotions and mood~\cite{morris1992functional}; theories state that despite being distinct mechanisms, they affect one another repeatedly and continuously. Nevertheless, hardly any research has been devoted towards computationally modelling the mood-emotion interplay for mood inference. 

\RS{Preliminary studies on mood recognition~\cite{sigal2010human, thrasher2011mood} infer mood via body posture and 3D pose data in a controlled setting. As a step towards \emph{in-the-wild} mood prediction,~\cite{narayana2022improve} uses deep learning to perform mood classification on affective videos. It observes that mood prediction improves on utilising emotion-change information. The premise in~\cite{narayana2022improve} is the existence of continuous emotion (valence) labels along with mood annotations during the training and testing phases. Reliance on continuous emotion labels represents a significant overhead, as these labels may not be available in real-world settings.} 

 \RS{This work is inspired by and extends the idea proposed in~\cite{narayana2022improve}, and obviates the need for ground-truth emotion labels by \textit{deducing} emotional change information and utilising it for mood inference. Our mood inference framework is illustrated in Fig.~\ref{fig:overview}. Specifically, emotion change is modelled in terms of 
 emotional (dis)similarity between a pair of video frames via a Siamese Network with contrastive loss. 
 
 We present mood inference results on the \textit{AffWild2} database~\cite{kollias2019expression}, where (a) video mood labels are derived as in~\cite{narayana2022improve}; (b) \emph{pseudo} emotion-change ($\Delta$) labels are derived via a pre-trained Siamese Network; (c) a 3-dimensional Convolutional Neural Network (3D-CNN) with a ResNet18 backbone and projection head is trained with mood labels; (d) a 3D-CNN with ResNet18 backbone and branched projection heads is trained with both mood and $\Delta$ labels, and (e) a Teacher-Student (TS) network~\cite{hinton2015distilling} is employed, where the teacher distills the privileged $\Delta$-specific knowledge to the student for mood inference.} 

 \RS{Consistent with~\cite{narayana2022improve}, we observe that mood prediction performance improves when emotion similarity information is incorporated, emphasising the prominence of short-term affect (emotion-change) for long-term affect (mood) inference. To summarise, the main contributions of this work are as follows:
\begin{enumerate}
\item We propose to infer mood employing \textit{emotional similarity}, which models emotion change between a pair of images. To this end, we train a 3D-CNN with two branches, which are respectively trained via video frames annotated with mood and $\Delta$ labels.
\item We weakly label the AffWild2~\cite{kollias2019expression} dataset, by generating $\Delta$ labels for video frame pairs using a pre-trained Siamese Network with contrastive loss. 
\item Assessing various models, we demonstrate that incorporating emotion-change ($\Delta$) information via emotion similarity benefits mood recognition and enhances mood prediction performance. Similar trends are observed in the experiments employing ground truth $\Delta_{GT}$ labels.
\item Through an ablation study, we verify the effectiveness of the various components of the proposed mood classification framework outlined in Fig.~\ref{fig:overview}.
\end{enumerate}}



\section{Related Work} 
\label{sec::related_work}

In this section, we present studies examining the affective phenomena of emotion and mood  (Sec.~\ref{sec::related_work_interplay}), the various affective databases available for affect inference, and machine learning studies examining mood inference (Sec.~\ref{sec::related_work_mood}). The motivation for this study given the literature context is presented in Sec.~\ref{subsec::relatedwork_novelty}.

\subsection{Emotion, Mood and their Interconnectedness} \label{sec::related_work_interplay}
While there are multiple definitions of emotion, the following characterisation appears to be consistent. Emotion is considered to be an episode of neurophysiological and cognitive change in response to an external or internal stimulus \cite{scherer1987toward}. The concepts of emotion and mood are distinguished based on the factors of duration, trigger, intensity, and behavioural impact~\cite{ekman1984expression}. The former are short-term, lasting for a few seconds and are elicited based on stimulus events. The level of response to the stimuli and the corresponding emotional expression is recognised to be of relatively high intensity~\cite{oatley2006understanding}. In contrast, moods are considered to be  enduring affective states, lasting for hours or even days without being instantiated by a stimulus. They are regarded to be diffuse with low levels of intensity~\cite{ekman1984expression}. 

The human physiological state and mind are both influenced by and reflective of mood, as it directly influences human health and well-being. Besides having an impact on evaluative judgements, mood also governs memory retrieval~\cite{wong2016mood}. Mood-congruency, which refers to the match between a person's mood and his/her thoughts~\cite{mayer1992mood}, is observed in~\cite{schmid2010mood}, where the authors examine mood effects on emotion recognition. Happy mood impedes the recognition of mood-incongruent sad emotions, while sad mood obstructs the recognition of happy emotions. The interplay between mood and emotion is described by the mood-emotion loop~\cite{wong2016mood}, a theory which proposes that mood and emotion are distinct mechanisms forming a loop,  and are reciprocally influencing one another. 

\subsection{Computational Studies on Mood} 
\label{sec::related_work_mood}
Most research on affective state inference has focussed on emotions, as opposed to mood. Likewise, the prevailing affective databases to aid behavioural and computational studies with various modes of data predominantly target emotions. 

\subsubsection{\textbf{Databases}} 
\label{subsec::relatedwork_databases}
Audio-visual databases, such as AFEW \cite{dhall2012collecting}, DECAF~\cite{Abadi15}, Ascertain~\cite{Subramanian18} \etc., comprise videos of emotional episodes with categorical emotion annotations, namely, happy, sad, fear, disgust, anger, surprise, and neutral. RECOLA~\cite{ringeval2013introducing}, AFEW-VA~\cite{kossaifi2017afew}, AffWild2~\cite{kollias2019expression}, \etc., manifest continuous emotion annotations of \emph{valence} (degree of pleasantness or unpleasantness) and \emph{arousal} (degree of excitement or calmness). These databases, designed by considering factors such as recording environment, duration, and annotation type, best suit emotion inference tasks. The test set of the AVEC 2013 challenge, annotated for the level of depression, is one of the few databases with mood annotations. However, depression, a mood disorder, is not a commonly observed mood state. EMMA~\cite{katsimerou2016crowdsourcing} is an acted video database recorded in a controlled setting with mood annotations. 

\subsubsection{\textbf{Computational Approaches}} \label{subsec::relatedwork_approaches}
In~\cite{thrasher2011mood}, the authors use body posture and head movement features to capture the affective state while listening music. A vertical position of the head is observed in a positive mood and a downward position otherwise. A 3-dimensional pose tracker is used in~\cite{sigal2010human} to infer physical attributes and the mood of the person by capturing walking motions. The authors of~\cite{katsimerou2015predicting} perform automatic mood recognition from recognised emotions, and show that clustered emotions in the valence-arousal space are better predictors of a single mood as compared to multiple moods within a video. 

Mood prediction using various 3D-CNNs are performed in~\cite{narayana2022improve} using the AFEW-VA~\cite{kossaifi2017afew} dataset. Utilising the valence annotations, the authors compute the valence differential to infer mood and demonstrate that incorporating valence change improves mood prediction performance. Although this study is a promising step towards automatic mood inference, only video clips of very short duration ($\approx$ 0.04 seconds) are considered, which may not adequately depict subject mood in the video. 

\subsection{Novelty of our Study} \label{subsec::relatedwork_novelty}
A thorough examination of the literature reveals the following: (a) While significant research has been conducted to infer emotions automatically, mood inference and modelling the mood-emotion interplay have been neglected from a computational perspective; (b) Existing affective databases are richly annotated for emotions, while labelled data for mood inference are sparse; and (c) Existing studies, which infer mood using emotions, require continuous valence annotations and consider clips of very short duration ($\approx$ 0.04 seconds) for this purpose. 

Differently, this study uses \textit{deduced} emotional similarity information in lieu of valence-differential ($\Delta$) labels, obviating the need for ground truth $\Delta$ annotations. This setting resembles real-world scenarios where valence annotations may not be accessible for inferring mood. This work proposes to (1) deduce emotion change ($\Delta$) labels using a Siamese Network trained with contrastive loss, and (2) incorporate $\Delta$ labels for inferring mood. In addition, ablation studies are performed to empirically examine the effect of different components of the proposed mood inference framework.


\section{Databases and Label Generation} 
\label{sec::databases_labels}
This section describes the databases considered in this study as well as the labelling procedure.

\subsection{AffWild2 Database} 
\label{subsec::databases_affwild2}
We consider AffWild2~\cite{kollias2019expression}, a publicly available affective video database with continuous dimensional (valence, arousal) and categorical emotion annotations for performing mood classification. AffWild2 comprises 564 \textit{in-the-wild} videos collected from \textit{YouTube}. There are a total of 2,816,832 frames with 455 subjects (277 male, 178 female). The videos are annotated by four experts for continuous valence and arousal values, and the average of the four raters is considered as the final rating, while three experts annotated the categorical emotion labels. The valence and arousal annotations are in the range $\left[-1, 1\right]$, and the emotional categories are \emph{happy}, \emph{sad}, \emph{disgust}, \emph{anger}, \emph{fear}, \emph{surprise}, \emph{neutral}, and \emph{other}. The frames with annotated values outside this range are discarded as suggested by the dataset providers. The partitioning of the database into training, validation and test sets is done in a subject-independent manner, so that every subject is present in one of the three partitions. This partitioning results in 341, 71 and 152 videos respectively in the training, validation and test sets. The validation set is utilized for evaluating our proposed approach, as the test set has not been released.

\subsection{Mood labels} 
\label{subsec::mood_labels}
Since current \textit{in-the-wild} affective video databases lack mood annotations, as the closest alternative, we utilize valence annotations to assign mood labels for each video in the AffWild2 database, as done in \cite{narayana2022improve}. The three mood categories considered are \emph{positive} (+1), \emph{negative} (-1), and \emph{neutral} (0). As mood denotes a long-term affective state~\cite{picard2000affective}, the most persistent valence value (maximum number of consecutive frames) is considered for assigning a mood label. Mood label is assigned to $-1$, $0$, or $+1$ if the valence values for maximum number of consecutive frames respectively lie in the range $[-1, -0.3)$, $[-0.3, 0.3]$, and $(0.3, 1]$. 

\subsection{AffectNet Database} 
\label{subsec::databases_affectnet}
In order to automatically generate the emotion change ($\Delta$) labels, we employ the \textit{AffectNet} \cite{mollahosseini2017affectnet} dataset to train a Siamese Network (described in Sec.~\ref{subsec::delta_labels}). AffectNet, an affective database curated for automatic facial emotion recognition tasks, comprises around 420,300 facial images captured under natural conditions. Twelve experts annotated the data with continuous valence and arousal values, and eight emotion categories (\emph{happy}, \emph{sad}, \emph{disgust}, \emph{anger}, \emph{fear}, \emph{surprise}, \emph{neutral}, and \emph{contempt}). Partitioning the dataset with the criteria of having 500 images for each of the nine emotion categories in the validation set, results in 287,151 and 4500 images in the training and test sets respectively. 

\subsection{Emotion Change ($\Delta$) Labels} 
\label{subsec::delta_labels}
\begin{figure}[t]
\centering
\includegraphics[width=0.9\linewidth]{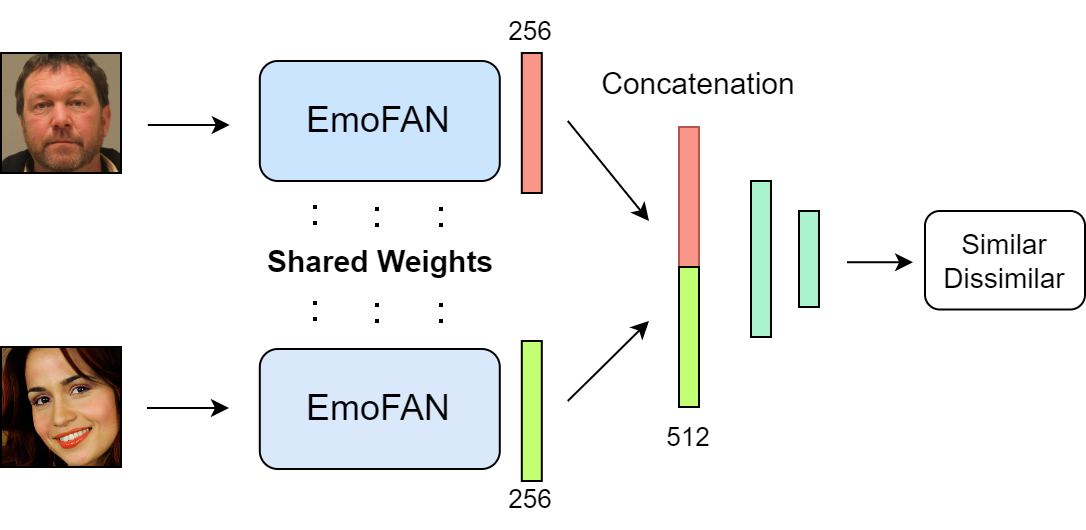}
\caption{Siamese network for modelling emotion-change ($\Delta$).}
\label{fig:siamese_net}
\end{figure}

This work seeks to eliminate the need for valence annotations (as proposed in~\cite{narayana2022improve}) to generate $\Delta$ labels. Differently, we propose to use a Siamese Network with contrastive loss to deduce emotion-change ($\Delta$) between video frames in terms of  similarity (little change in emotion) or dissimilarity (significant change in emotion). A Siamese network is a neural network that discriminates if a pair of input data samples are similar or dissimilar \cite{bromley1993signature}. Multiple emotion-related tasks have employed a Siamese network, and observed promising performance~\cite{lian2018speech, sabri2018facial}. We use contrastive loss in the Siamese network, which learns a similarity metric that minimises the distance between similar image pairs, while maximising the distance between dissimilar pairs. Distance between the pairs is compared to a \textit{margin} value, and the contrastive loss function enforces a smaller distance between 
similar pairs, and a larger distance between dissimilar pairs.

\subsubsection{\textbf{Siamese Network}}
In this study, \emph{similarity} refers to little or no change in the emotional facial expression between a pair of frames. Fig.~\ref{fig:siamese_net} shows the architecture of the Siamese network, which comprises identical sub-networks for classifying the input frames as \emph{similar} or \emph{dissimilar}. Each sub-network involves an EmoFAN~\cite{toisoul2021estimation} network as the encoder, $E(\cdot)$, which maps the input image to a vector. We employ a pre-trained EmoFAN model as it has demonstrated high performance in emotion recognition tasks. For two images $x_1$ and $x_2$ we obtain, $v_1 = E(x_1) \in \mathcal{R^{D_{E}}}$ and $v_2 = E(x_2) \in \mathcal{R^{D_{E}}}$, where $\mathcal{D_{E}} = 256$. The embeddings $v_1$ and $v_2$ are concatenated, $v = v_1 \mathbin\Vert v_2 \in \mathcal{R^{D_{C}}}$, where $\mathcal{D_{C}} = 512$. $v$ is fed into a projection head $P(\cdot)$, which maps it to a vector, $u = P(v) \in \mathcal{R^{D_{P}}}$, where $\mathcal{D_{P}} = 2$. $P(\cdot)$ is a Multi-Layer Perceptron (MLP) with three fully connected (fc) layers, comprising 256, 128, and 2 neurons, respectively. The neurons in the last fc layer refer to the two classes, \emph{similar} and \emph{dissimilar}. The inputs to the fc layers are normalised with zero mean and unit variance, before feeding to the ReLU activation.

\subsubsection{\textbf{Contrastive Loss}}
As opposed to using the cross-entropy loss, $\mathcal{L_{B}}$, alone for binary classification (similar/dissimilar),  we additionally consider using the contrastive loss, $\mathcal{L_{C}}$, given by, 
\begin{equation}
\vspace{-2mm}
\mathcal{L_{C}} = \frac{1}{N} \sum_{i=1}^N y_{i}(1 - d_{i}) + (1 - y_{i}) \max(0, d_{i} - m)
\end{equation}
where $N$ is the batch size, $y_i$ is the label indicating whether the two input samples are similar (1) or dissimilar (0), $d_i$ is the cosine distance between the embeddings $v_1$ and $v_2$, and $m$ is the margin. The total loss is given by, 
\begin{equation}
\mathcal{L_{T}} = \lambda \mathcal{L_{B}} + (1 - \lambda) \mathcal{L_{C}}
\end{equation}
where $\lambda$ is a training hyperparameter.

\subsubsection{\textbf{Deducing the $\Delta$ Label}}
The Siamese network is trained on the images in the AffectNet database. The groundtruth labels for the Siamese network $y_i$ are derived as 1 if the emotion categories of the input pair are the same, 0 otherwise. The network is trained for 40 epochs, using the Adam optimiser with the learning rate decreased by a factor of 10 for every 10 epochs, with the initial learning rate as 0.0001. The batch size is set to 64, with the dropout rate as 0.3, margin $m$ in the contrastive loss set to 0.25, and $\lambda$ set to 0.5.

The Siamese network achieves an accuracy of 68\%. This model is used to generate $\Delta$ labels for the AffWild2 video frame pairs.   

\subsection{Generating input samples for mood inference} 
\label{subsec::input_samples}
\begin{figure*}[t]
\centering
\includegraphics[width=\textwidth]{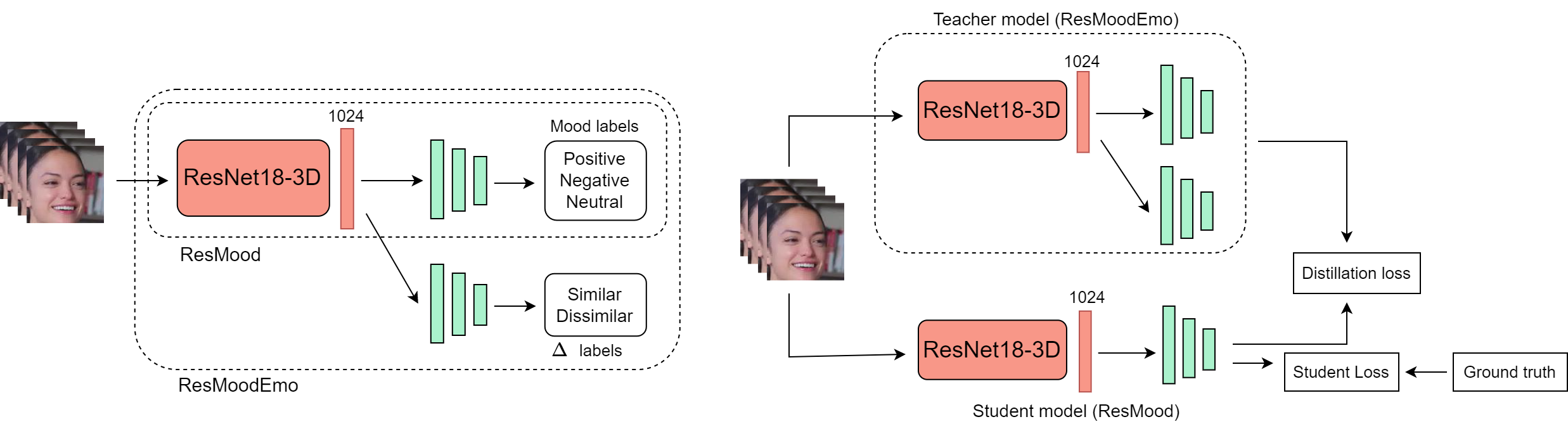}
\caption{(Left) The architecture of ResMood is shown within the inner dashed rectangle. The outer dashed rectangle represents ResMoodEmo. (Right) The architecture of the TS-Net. (Best viewed in colour).}\vspace{-2mm}
\label{fig:resmoodemo_tsnet}
\end{figure*}

The AffWild2 dataset comprises videos of long duration, with an average length of $\approx$ 3 minutes (minimum duration of 0.03 minutes, maximum duration of 26.22 minutes). From each video, using a sliding window approach, we generate clips with a stride $s$, where each clip is a collection of sampled frames. Each clip $c$ is of temporal length $t$, which refers to the duration of the clip (number of frames). Constructing clips by including all the frames in $c$ increases the computational load and time substantially. Hence, to address this computational impediment, we significantly reduce the number of frames in $c$, and sample $n$ frames at equal intervals of time. Clips generated from each video contain frames from the parent video alone, and no frames from other videos. 

$c$ is assigned the mood label of its parent video, implying all clips generated from a source video are assigned the same mood label. To generate the $\Delta$ label for $c$, the first frame and the last frame of $c$ are fed to the trained Siamese Network (described in Sec.~\ref{subsec::databases_affectnet}). The model returns 0 or 1 as the $\Delta$ label, by evaluating the dissimilarity or similarity of the emotion displayed in this pair of frames. Hence, each clip is associated with a mood label $\in$ \{0, +1, -1\}, and a $\Delta$ label $\in$ \{0, 1\}.


\section{Mood Classification Approach} 
\label{sec::approaches}


The capability of 3D-CNNs to capture the temporal dependencies in the input data, along with spatial information, has resulted in their extensive usage. ResNet18-3D~\cite{lu2018multiple} (R3D), a 3D variant of the ResNet architecture, is commonly used as a backbone network in many 3D-CNN architectures for facial emotion inference ~\cite{kim2022emotion, zhou2019exploring}. Leveraging the interplay between mood and emotions, we utilise the mood and $\Delta$ labels to perform mood classification. The various models used in this study are described as follows.

\subsection{ResMood} 
\label{subsec::approaches_resmood}
Fig.~\ref{fig:resmoodemo_tsnet} (left) shows ResMood, a model trained with mood labels alone, which consists of a ResNet18-3D, $R3D(\cdot)$, as the backbone and a projection head, $P(\cdot)$. The backbone maps each input sample $x$ to a representation vector, $v = R3D(x) \in \mathcal{R^{D_B}}$, where $\mathcal{D_B} = 1024$. The projection head $P(\cdot)$ further maps $v$ to a vector $z = P(v) \in \mathcal{R^{D_{P}}}$, where $\mathcal{D_P} = 3$. $P(\cdot)$ is instantiated as a Multi-Layer Perceptron (MLP), with three fully-connected (fc) layers comprising 512, 256, and 3 neurons, respectively. The 3 neurons in the last fc layer denote the three mood classes, \emph{positive}, \emph{negative}, and \emph{neutral}. The inputs to each layer in the MLP are normalised batch-wise with zero mean and unit variance before feeding them to the ReLU activation function.

\subsection{ResMoodEmo}
\label{subsec::approaches_resmoodemo}
Fig.~\ref{fig:resmoodemo_tsnet} (left) shows ResMoodEmo, a model for performing mood classification trained with both mood and $\Delta$ labels. Distinct from ResMood, ResMoodEmo is composed of $R3D(\cdot)$ as the backbone, and two projection heads $P_{M}(\cdot)$ and $P_{\Delta}(\cdot)$, branching out for mood and $\Delta$ classification, respectively. $R3D(\cdot)$ maps an input $x$ to a vector $v = R3D(x) \in \mathcal{R^{D_B}}$, where $\mathcal{D_B} = 1024$. Further, the projection $P_{M}(\cdot)$ maps $v$ to a vector $z_1 = P_{M}(v) \in \mathcal{R^{D_{M}}}$, and $P_{\Delta}(\cdot)$ maps $v$ to a vector $z_2 = P_{M}(v) \in \mathcal{R^{D_{\Updelta}}}$, where $\mathcal{D_M} = 3$ and $\mathcal{D_{\Updelta}} = 2$, respectively. $P_{M}(\cdot)$ and $P_{\Delta}(\cdot)$ are both configured as MLPs with three fc layers, but differing in the number of neurons in the last layer. As $P_{M}(\cdot)$ is the branch used for mood classification, the last fc layer has 3 neurons for classifying the three mood classes \emph{positive}, \emph{negative}, and \emph{neutral}, whereas $P_{\Delta}(\cdot)$ used for classifying $\Delta$ labels has two neurons denoting the \emph{similar} and \emph{dissimilar} classes. In both projection heads, the inputs to each layer are normalised with zero mean and unit variance prior to being input to the ReLU activation function. Compared to ResMood, ResMoodEmo has an additional branch after the $R3D(\cdot)$ to incorporate the emotion change ($\Delta$) information. The losses of each branch $\mathcal{L_M}$ and $\mathcal{L_{\Updelta}}$ are summed up and the cumulative loss $\mathcal{L} = \mathcal{L_M} + \mathcal{L_{\Updelta}}$ is optimised. 

\vspace{-2mm}
\subsection{Teacher-Student Network} 
\label{subsec::approaches_ts_net}
\vspace{-2mm}
Similar to~\cite{narayana2022improve}, we employ knowledge distillation~\cite{hinton2015distilling}, a technique used to transfer knowledge from a larger (teacher) model to a smaller (student) model. In this method, the goal is to train the student model to mimic the output probabilities of the teacher model, in addition to the predicting the true labels. Fig.~\ref{fig:resmoodemo_tsnet} (right) presents the Teacher-Student Network (TS-Net), where ResMoodEmo is used as the teacher model (see Sec.~\ref{subsec::approaches_resmoodemo}), and ResMood is used as the student model (see Sec.~\ref{subsec::approaches_resmood}). The teacher model, trained with both mood and $\Delta$ labels, distills knowledge to the student, which is only trained with mood labels. Since the performance of the student model alone is evaluated, $\Delta$ labels are not utilised during the testing phase. The SoftMax layer of the student model has a hyper-parameter called \emph{temperature} ($T$), which regulates the softness of the output class probabilities. Using low temperature values produces a sharper probability distribution, facilitating the student to focus on the relative differences in the probabilities of the classes. A weighted sum of the distillation loss, $\mathcal{L_D}$, measuring the difference between the outputs of the teacher and student models, and the student loss, $\mathcal{L_S}$, a typical supervised loss is optimised in the TS-Net, $\mathcal{L} = \alpha \mathcal{L_S} + (1 - \alpha) \mathcal{L_D}$, where $\alpha$ is a training hyperparameter.

\subsection{Implementation details} \label{subsec::approaches_implementation}
All experiments are based on using the open-source library PyTorch. The models are trained on Nvidia GeForce RTX 3090 GPU with 24GB memory. We use the videos with cropped and aligned faces provided in the AffWild2 database. To generate the input samples (see Sec.~\ref{subsec::input_samples}), we set the temporal length $t = 100$, with the number of frames in each sample $n = 5$, and the stride $s = 3$. The models ResMood, ResMoodEmo, and the TS-Net are trained using the Adam optimiser with the learning rate reduced by a factor of 10 for every 10 epochs, and the base learning rate set to 0.0001. The models are trained for 30 epochs with a batch size of 128 and the dropout rate is 0.5. In the TS-Net, we validated with the temperature values $\in$ $\{3, 5, 7\}$ and $\alpha \in \{0.05, 0.1, 0.15, 0.2\}$, as shown in Table \ref{tab:ablation_temp_alpha}. 


\section{Results and Discussion} 
\label{sec::results}
Due to an imbalance in mood classes in the test set, we use weighted F1-score as the performance evaluation metric in all our experiments. Table~\ref{tab:mood_delta} shows the results of ResMood, ResMoodEmo, and TS-Net. While ResMood is trained with mood labels alone, ResMoodEmo is trained with both mood and $\Delta$ labels. In the TS-net, ResMoodEmo is the teacher model, and ResMood is the student model, implying that the teacher is pre-trained with both mood and $\Delta$ labels, while the student is trained with mood labels alone. 

ResMoodEmo yields a higher F-score as compared to ResMood, indicating that the ResMoodEmo learning temporal short-term emotion changes, better predicts mood than ResMood, which only employs mood labels. Without resorting to the valence differential for gathering emotion change information as done in \cite{narayana2022improve}, training a Siamese Network with contrastive loss, and generating the $\Delta$ labels results in a competitive performance in mood prediction. Further, by using the emotion change labels, we are capturing variations over a short duration simultaneously for characterising mood. The results indicate that (local) emotion variations contribute towards understanding the (global) mood.  

A similar trend is observed in the TS-Net, as it yields a higher F-score than the ResMood model. This shows that the teacher, possessing the privileged knowledge concerning emotion change, is able to effectively distil knowledge to the student (ResMood) during the training phase. With the $\Delta$ labels being implicitly given as soft labels from ResMoodEmo, the performance of the student increases, as compared to the standalone ResMood model. Cumulatively, these results show that using the pseudo-emotion-change information enhances the mood prediction performance.

For the Siamese Network trained on the AffectNet dataset, generating effective $\Delta$ labels from AffWild2 is crucial. The obtained results show that the $\Delta$ labels generated are reliable as they improve mood prediction performance; the investigation of an optimal architecture for the Siamese network is left to future work. Overall, our results confirm that despite not using the valence differential labels to denote emotion changes, performing weak supervision using the $\Delta$ labels in ResMoodEmo and TS-Net improves mood prediction performance. 

{\renewcommand{\arraystretch}{1.2}%
\begin{table}[t]
\caption{Performance results (weighted F-score) of the models.}
\centering
\label{tab:mood_delta}
\resizebox{0.82\linewidth}{!}{%
\begin{tabular}{|c|c|c|}
\hline
Model & Train labels & F-score \\ 
\hline\hline
ResMood & Mood & 0.65 \\
ResMoodEmo & Mood and $\Delta$ & \textbf{0.78} \\
TS-Net & Mood (Student) & \textbf{0.78} \\ \hline
\end{tabular}%
}\vspace{-3mm}
\end{table}

\subsection{Ablation Studies} \label{subsec::ablation}

{\renewcommand{\arraystretch}{1.2}%
\begin{table}[t]
\caption{Ablation study results with $\Delta_{GT}$ labels}
\centering
\label{tab:emo_ground_truth}
\resizebox{0.82\linewidth}{!}{%
\begin{tabular}{|c|c|c|}
\hline
Model & Train labels & F-score \\ 
\hline\hline
ResMood & Mood & 0.63 \\
ResMoodEmo & Mood and $\Delta_{GT}$ & 0.73 \\
TS-Net & Mood (Student) & 0.66 \\ \hline
\end{tabular}%
}\vspace{-3mm}
\end{table}

{\renewcommand{\arraystretch}{1.2}%
\begin{table*}[t]
\caption{Ablation study varying number of frames ($n$) in the input samples. Best results obtained are highlighted in bold.}
\centering
\label{tab:ablation_num_frames}
\resizebox{0.85\textwidth}{!}{%
\begin{tabular}{|c|c|cc|cc|}
\hline
\multirow{2}{*}{Number of frames} & ResMood & \multicolumn{2}{c|}{ResMoodEmo} & \multicolumn{2}{c|}{TS-Net} \\ \cline{2-6} 
 & F-score & \multicolumn{1}{c|}{F-score} & \% increase to ResMood & \multicolumn{1}{c|}{F-score} & \% increase to ResMood\\ \hline\hline
3 & 0.67 & \multicolumn{1}{c|}{0.68} & +1.49 & \multicolumn{1}{c|}{0.67} & 0 \\
5 & 0.65 & \multicolumn{1}{c|}{\textbf{0.78}} & \textbf{+20} & \multicolumn{1}{c|}{\textbf{0.78}} & \textbf{+20} \\
7 & 0.68 & \multicolumn{1}{c|}{0.72} & +5.88 & \multicolumn{1}{c|}{0.65} & -4.41 \\
9 & 0.70 & \multicolumn{1}{c|}{0.63} & -10 & \multicolumn{1}{c|}{0.52} & -25.71 \\ \hline
\end{tabular}%
}\vspace{-3mm}
\end{table*}}

{\renewcommand{\arraystretch}{1.3}%
\begin{table*}[]
\caption{Ablation study results for various temporal lengths ($t$) of the input sample.}
\centering
\label{tab:ablation_temp_length}
\resizebox{0.85\textwidth}{!}{%
\begin{tabular}{|c|c|cc|cc|}
\hline
\multirow{2}{*}{Temporal length} & ResMood & \multicolumn{2}{c|}{ResMoodEmo} & \multicolumn{2}{c|}{TS-Net} \\ \cline{2-6} 
 & F-score & \multicolumn{1}{c|}{F-score} & \% increase to ResMood & \multicolumn{1}{c|}{F-score} & \% increase to ResMood\\ 
 \hline\hline
50 & 0.67 & \multicolumn{1}{c|}{0.65} & -2.99 & \multicolumn{1}{c|}{0.66} & -1.49 \\
100 & 0.65 & \multicolumn{1}{c|}{\textbf{0.78}} & \textbf{+20} & \multicolumn{1}{c|}{\textbf{0.78}} & \textbf{+20} \\
150 & 0.69 & \multicolumn{1}{c|}{0.65} & -5.80 & \multicolumn{1}{c|}{0.64} & -7.25 \\
200 & 0.65 & \multicolumn{1}{c|}{0.64} & -1.54 & \multicolumn{1}{c|}{0.64} & -1.54 \\ \hline
\end{tabular}%
}\vspace{-3mm}
\end{table*}}

{\renewcommand{\arraystretch}{1.3}%
\begin{table*}[]
\caption{Ablation study results for various backbone architectures.}
\centering
\label{tab:ablation_backbone}
\resizebox{0.81\textwidth}{!}{%
\begin{tabular}{|c|c|cc|cc|}
\hline
\multirow{2}{*}{Backbone} & ResMood & \multicolumn{2}{c|}{ResMoodEmo} & \multicolumn{2}{c|}{TS-Net} \\ \cline{2-6} 
 & F-score & \multicolumn{1}{c|}{F-score} & \% increase to ResMood& \multicolumn{1}{c|}{F-score} & \% increase to ResMood\\ 
 \hline\hline
ResNet18 & 0.65 & \multicolumn{1}{c|}{\textbf{0.78}} & \textbf{+20} & \multicolumn{1}{c|}{\textbf{0.78}} & \textbf{+20} \\
ResNet34 & 0.66 & \multicolumn{1}{c|}{0.70} & +6.06 & \multicolumn{1}{c|}{0.66} & 0 \\
ResNet50 & 0.68 & \multicolumn{1}{c|}{0.75} & +7 & \multicolumn{1}{c|}{0.75} & +7 \\ \hline
\end{tabular}%
}\vspace{-3mm}
\end{table*}}

To corroborate the above findings, we perform the following ablation studies and examine the effectiveness of various components of our approach.

\subsubsection{\textbf{Using Ground Truth Emotion-change ($\Delta_{GT}$) Labels}}
The increase in the F-score using ResMoodEmo as compared to ResMood, could be attributed to the contribution of the $\Delta$ labels for mood inference or the implicit efficiency of the Siamese Network. For better comprehension of the results, we use the eight emotion labels available for each video frame in the AffWild2 dataset to obtain $\Delta_{GT}$ (\emph{similar} or \emph{dissimilar}) labels. However, since not all videos have the categorical emotion annotations, the total number of samples reduced from 191,552 to 53,026. These samples are obtained as described in Sec.~\ref{subsec::input_samples}. For each video sample, there is an assigned mood label and $\Delta_{GT}$ label (similarity between first frame and last frame). The results of ResMood and ResMoodEmo are shown in Table \ref{tab:emo_ground_truth}. It is noteworthy that a similar trend to Table \ref{tab:mood_delta} is observed here. This indicates that, (a) $\Delta$ labels generated by the Siamese Network are effective, as they result in a competitive performance, (b) emotion information positively contributes in mood inference in a dataset-agnostic manner ($\Delta$ labels are deduced from the Siamese Network trained using AffectNet, whereas the $\Delta_{GT}$ labels are obtained from AffWild2.), and (c) achieving a comparable result without using the $\Delta_{GT}$ highlights the robustness of the proposed mood inference approach. 

\subsubsection{\textbf{Number of Frames in the Input Video Sample}}
Table \ref{tab:ablation_num_frames} shows the results of varying the number of frames in the input samples, while fixing the temporal length ($t$) to 100. Temporal information plays a crucial role in examining mood, and variation in the number of frames in each sample clarifies if increasing the information provided to the model facilitates mood inference. The performance of ResMoodEmo increases when the number of frames are set to 3, 5, or 7, but decreases when the number of frames is increased to 9. Using the TS-Net, when the input samples have 3 frames, no change is observed, while with 7 and 9 frames, the performance reduces. Overall, the comparison shows that 5 frames in the input sample maximally increases mood prediction performance.

\subsubsection{\textbf{Temporal Length ($t)$ of the Sample}}
Table \ref{tab:ablation_num_frames} presents the results of varying $t$ of the input samples while fixing the number of frames ($n$) in each sample to be 5. Since mood is an enduring affect, it is important to consider long sequences of data for automatic mood inference. For $t = 50$, and $t = 150$, the F-score obtained with ResMood increases, while the F-score remains the same for $t = 200$, as compared to using $t = 100$. Although varying $t$ results in a comparable F-score with ResMood, it decreases in ResMoodEmo and TS-Net with lengths of 50, 150, and 200, as shown in the respective \emph{\% increase to ResMood} columns of Table \ref{tab:ablation_temp_length}. The highest F-score with ResMood and the least F-score with ResMoodEmo is observed using $t = 150$. Since emotion is a short-term affect, observing the changes in the emotion over longer sequence of time causes a detrimental effect on mood inference. 

\subsubsection{\textbf{Varying the Backbone Architecture}}
Table \ref{tab:ablation_backbone} reports the results when the backbone architecture in the models is changed. The F-score for ResMood increases slightly as the depth of the ResNet increases. With ResNet18 and ResNet50, a general trend of increase in the mood prediction performance using ResMoodEmo and TS-Net is observed. Using ResNet34, an increase in F-score is observed with ResMoodEmo as compared to ResMood, but with TS-Net, the F-score remains the same. ResNet18, a lighter architecture as compared to its counterparts, results in the maximum F-score for ResMoodEmo and TS-Net, and largest \% increase from ResMood.

\subsubsection{\textbf{Temperature and $\alpha$}}
The results with varying temperature ($T$) and $\alpha$ values are shown in Table \ref{tab:ablation_temp_alpha}. For varying $\alpha$, as $T$ increases, the F-score reduces. This is due to the fact that high values of temperature soften the output class probabilities, while with low temperature values, the relative differences in the probabilities are captured. For $T = 5$ and $T = 7$, the F-score either remains the same or increases as $\alpha$ increases, while for $T = 3$, no general trend is observed. As described in Sec.~\ref{subsec::approaches_ts_net}, $\alpha$ is the weight of the student loss function, indicating that lower values of $\alpha$ imply a higher weighting for the distillation loss, enabling the student model to get closer to the teacher model.

Overall, the ablation study results confirm that the generated $\Delta$ labels are an effective alternative to $\Delta_{GT}$ and produce a comparable effect, as incorporating these labels improves mood prediction performance.

{\renewcommand{\arraystretch}{1.3}%
\begin{table}[]
\caption{Ablation study results for the TS-Net with various temperature and alpha values.}
\centering
\label{tab:ablation_temp_alpha}
\resizebox{0.65\linewidth}{!}{%
\begin{tabular}{|c|c|c|c|c|}
\hline
T/$\alpha$ & 0.05 & 0.1 & 0.15 & 0.2 \\ \hline\hline
3 & \textbf{0.78} & 0.72 & 0.69 & 0.72 \\
5 & 0.64 & 0.64 & 0.68 & 0.69 \\
7 & 0.64 & 0.66 & 0.66 & 0.71 \\ \hline
\end{tabular}%
}\vspace{-2mm}
\end{table}}

\section{Conclusion} 
\label{sec::conclusion}
\vspace{-1mm}
The aim of this study is to examine mood from a computational perspective by incorporating emotion similarity information. Different from prior studies, without using the valence differential, this study proposes to use emotion change information by employing a metric learning approach. To this end, a Siamese network is trained using the AffectNet database and the trained model is used to generate pseudo-$\Delta$ labels for a pair of frames in the AffWild2 database. For mood classification, we employ ResMood, a model trained with mood labels alone, ResMoodEmo trained with mood labels and $\Delta$ labels, and TS-Net, a teacher-student network with ResMoodEmo as the teacher to distil knowledge to ResMood. Higher F-scores are observed with models trained with both mood and $\Delta$ labels as compared to models trained with mood labels alone. This indicates that the emotion change labels are generated effectively and contribute positively to the mood prediction performance. Our claim is further confirmed by similar trends when performing corresponding experiments employing $\Delta_{GT}$ labels.

\section*{Ethical Impact Statement}
\vspace{-1mm}
This study aims at examining mood from a computational perspective by using emotional similarity information. The data for the study reuses publicly available databases, AffectNet and AffWild2, to conduct computational modeling experiments. This study is designed towards answering a theoretical question regarding the interaction between mood and emotion. While facial information is revealed from images and videos in the databases, we neither use identity-specific information, nor base our claims on a specific religion, race or gender. The proposed framework is non-obtrusive, using the images, and videos present in the databases. One of the most crucial applications of this framework is in healthcare, to detect early signs of mood disorders such as depression, and monitoring the mood of the patients by observing their emotional patterns. Other application include education, gaming technology, marketing, etc \cite{picard2000affective}. 

Although we aim at developing robust mood inference technology, as with any other affect recognition system, there could be potential ethical concerns. Mood inference could reveal sensitive information about an individual's mental state, and could be used against the person. Mood detection system could be also used inappropriately to influence or manipulate individuals' behavior or emotions. Additionally, the use of mood detection in contexts such as employment could lead to discrimination or bias. Finally, we acknowledge that there could be intrinsic bias, as we train our models on the databases which may be biased towards facial expressions of individuals from a specific location/culture.

\bibliography{references.bib}{}
\bibliographystyle{IEEEtran}

\end{document}